\documentclass[prd,aps,twocolumn,preprintnumbers,superscriptaddress,showpacs,floatfix,nofootinbib]{revtex4-1}
\pdfoutput=1

\usepackage{tikz}
\usepackage{booktabs}
\usepackage[toc,page]{appendix}
\usepackage{amsmath,bbm,array,amsfonts,float,mathtools,amsthm}
\usepackage{multirow}
\usepackage{caption}
\usepackage{microtype}
\usepackage{hyperref}
\usepackage[toc,page]{appendix}
\usepackage[utf8]{inputenc}
\usepackage[autostyle=true]{csquotes}
\usepackage{todonotes}
\usepackage{graphicx}
\usepackage{environ}
\usepackage[framemethod=Tikz]{mdframed}

\usetikzlibrary{decorations.pathmorphing}
\usetikzlibrary{decorations.pathreplacing}
\usetikzlibrary{decorations.markings}
\usetikzlibrary{shapes, shapes.geometric, shapes.symbols, shapes.arrows, shapes.multipart, shapes.callouts, shapes.misc}
\usetikzlibrary{shadows,fit,trees,calc,patterns,arrows}
\tikzset{snake it/.style={decorate, decoration=snake, segment length=2mm}}
\tikzset{7brane/.style={circle, draw=black, fill=black,ultra thick,inner sep=1.5 pt, minimum size=1 pt,}, c/.default={4pt}}
\tikzset{cross/.style={cross out, draw=black,ultra thick, minimum size=2*(#1-\pgflinewidth), inner sep=0pt, outer sep=0pt}, cross/.default={5pt}}

\newcommand{\ZZ}{\mathbb{Z}}

\numberwithin{equation}{section}
\numberwithin{figure}{section}
\numberwithin{table}{section}
\hypersetup{
    colorlinks,
    linkcolor={red!50!black},
    citecolor={blue!50!black},
    urlcolor={blue!80!black}
}

\begin{document}

\title{Brain Webs for Brane Webs}

\author{Guillermo Arias-Tamargo}
\email{guillermo.arias.tam@gmail.com}
\affiliation{Department of Physics, Universidad de Oviedo, C/ Federico Garc\'{i}a Lorca 18, 33007, Oviedo, Spain}
\affiliation{Instituto Universitario de Ciencias y Tecnolog\'{i}as Espaciales de Asturias (ICTEA)
C/ de la Independencia 13, 33004 Oviedo, Spain}
\author{Yang-Hui He}
\email{hey@maths.ox.ac.uk}
\affiliation{Department of Mathematics, City, University of London, EC1V 0HB, UK}
\affiliation{London Institute for Mathematical Sciences, Royal Institution, London W1S 4BS, UK}
\affiliation{Merton College, University of Oxford, OX1 4JD, UK}
\affiliation{School of Physics, NanKai University, Tianjin, 300071, P.R. China}
\author{Elli Heyes}
\email{elli.heyes@city.ac.uk}
\affiliation{Department of Mathematics, City, University of London, EC1V 0HB, UK}
\affiliation{London Institute for Mathematical Sciences, Royal Institution, London W1S 4BS, UK}
\author{Edward Hirst}
\email{edward.hirst@city.ac.uk}
\affiliation{Department of Mathematics, City, University of London, EC1V 0HB, UK}
\affiliation{London Institute for Mathematical Sciences, Royal Institution, London W1S 4BS, UK}
\author{Diego Rodriguez-Gomez}
\email{d.rodriguez.gomez@uniovi.es}
\affiliation{Department of Physics, Universidad de Oviedo, C/ Federico Garc\'{i}a Lorca 18, 33007, Oviedo, Spain}
\affiliation{Instituto Universitario de Ciencias y Tecnolog\'{i}as Espaciales de Asturias (ICTEA)
C/ de la Independencia 13, 33004 Oviedo, Spain}

\preprint{LIMS-2022-08}

\begin{abstract}
Abstract: We propose a new technique for classifying 5d Superconformal Field Theories arising from brane webs in Type IIB String Theory, using technology from Machine Learning to identify different webs giving rise to the same theory. We concentrate on webs with three external legs, for which the problem is analogous to that of classifying sets of 7-branes. Training a Siamese Neural Network to determine equivalence between any two brane webs shows an improved performance when webs are considered equivalent under a weaker set of conditions. This therefore suggests that the conjectured classification of 7-brane sets is not complete. 
\end{abstract}

\maketitle

\section{Introduction \& Summary}
\label{sec:intro}

Interacting UV complete Quantum Field Theories (QFTs) in $d>4$ are notoriously hard to construct, yet very interesting for a number of reasons, ranging from those more theoretical --the study of the structure of QFT across dimensions, which has proven to be a vantage perspective on strong coupling physics and dualities-- to those more applied --such as extra-dimensional models for our world, in many occasions string-inspired. 
Concentrating on the 5d case, one may try to construct 5d fixed point theories through an $\epsilon$-expansion of asymptotically free gauge theories in $4+\epsilon$ dimensions. Unfortunately, it is unclear whether this approach can be extended all the way to $d=5$ by setting $\epsilon=1$. 

The situation is, however, very different for supersymmetric 5d QFTs. In \cite{Seiberg:1996bd}, using String Theory, Seiberg argued that $SU(2)$ gauge theories with $N_f<8$ flavor hypermultiplets can be regarded as a certain mass-deformation of strongly coupled non-lagrangian UV Super-Conformal Field Theories (SCFTs) with $E_{N_f+1}$ global symmetry, thus paving the road to the study of 5d QFTs by String Theory methods.
Now, 5d SCFTs appear in different guises in String/M theory. One avatar is through compactifications on Calabi-Yau spaces \cite{Morrison:1996xf}. Another is as low energy descriptions of systems of 5-branes in IIB String Theory \cite{Aharony:1997ju,Aharony:1997bh}, a scenario on which we shall focus here. However, it is not known whether these approaches allow the construction of all possible 5d SCFTs, nor whether a given SCFT admits a description in both languages. 

Systems of 5-branes overlapping in 4+1 dimensions are completely described by their arrangement on a transverse 2d plane where each 5-brane looks like a segment. As these segments can meet and recombine, they form a ``web" in the plane. Moreover, we will assume that the external 5-brane legs of the web end on a suitable 7-brane --which in the plane of the web looks like a point--, so that the web ``hangs" from a 7-brane set. In this language a 5d SCFT is simply a 5-brane web such that all external legs meet at a point. This immediately suggests that it must be possible to construct and classify 5d SCFTs --or at least ones arising from branes-- in a very simple way by simply listing all possible consistent brane webs. This classification program is of obvious interest, and would complement the work initiated in \cite{Jefferson:2017ahm,Bhardwaj:2020gyu,Apruzzi:2019opn,Apruzzi:2019enx,Bhardwaj:2020ruf,Bhardwaj:2020avz}.

Nevertheless, one quickly realizes that, at least stated as above, the set of all consistent brane webs is hugely redundant, coming from two sources. First, the $SL(2,\mathbb{Z})$ duality of Type IIB String Theory translates into a global equivalence. Second, since the length of the external legs is not a parameter in the 5d QFT, one may imagine reducing this size (i.e., moving the external 7-brane along the corresponding leg) and eventually crossing it to the other side. Since 7-branes come with a branch cut for the axiodilaton, rotating it so that it does not cross the web induces an $SL(2,\mathbb{Z})$ transformation of part of the web as well as the creation/annihilation of the appropriate number of 5-branes. This whole process --to which we will refer as a Hanany-Witten (HW) move-- results in the equivalence of otherwise seemingly different webs \cite{Hanany:1996ie}. 

Thus, coming back to our classification problem, the set of all possible 5d SCFTs arising from brane webs is the set of all consistent branewebs modulo the equivalence relation set by (a) $SL(2,\mathbb{Z})$ duality and (b) HW moves.
In principle, implementing this program is straightforward. For instance, one could imagine charting the space of 5d SCFTs by first fixing the number of external legs, then drawing all possible webs and lastly keeping only one representative of each equivalence class under (a) and (b). However, in practice this is very hard, mostly due to the difficulty in implementing the equivalence relations. Thus, our classification problem seems insurmountable, as equivalent pairs need to be identified largely through trial and error. 
It is at this point where new Machine-Learning (ML) techniques, which came into String Theory in \cite{He:2017aed,He:2017set,Krefl:2017yox,Ruehle:2017mzq,Carifio:2017bov}, may come to the rescue and provide an invaluable tool to accomplish this classification of webs.

Interestingly, since brane webs end on 7-branes, our problem is closely related to that of classifying sets of 7-branes (in fact, for the case of webs with three legs studied in this paper, both problems are equivalent). In turn, this problem, affected by the same ambiguities due to $SL(2,\mathbb{Z})$ and HW moves, was studied long ago in \cite{DeWolfe:1998eu}. In that reference it is conjectured that sets of 7-branes are indeed fully characterized by a certain set of quantities (reviewed in Section \ref{sec:webdata}). However, it turns out that this conjecture is not quite correct, as it is possible to construct sets of 7-branes with the same classifiers which are nevertheless not equivalent (we present explicit examples in Section \ref{sec:webdata}). Our novel approach through ML may therefore shed further light on the classification of 7-branes as well.

ML, as a subfield of artificial intelligence, centres itself on the development of predominantly statistical tools to recognise and study patterns in large datasets. Neural Networks (NNs) are a primary tool within supervised ML, whose application on labelled data acts as a non-linear function fitting to map inputs to outputs, both represented as tensors over $\mathbb{Q}$ using decimals. In recent years the advancement of computational power has played perfectly into the hands of these many-parameter techniques, leading to a programme of application of these tools to datasets arising in theoretical physics \cite{Ashmore:2019wzb,Douglas:2020hpv,Jejjala:2020wcc,Anderson:2020hux,Halverson:2019tkf,Brodie:2019dfx,Krippendorf:2021lee,Bao:2020nbi,Bao:2021auj,Chen:2020dxg,Larfors:2021pbb} and the relevant mathematics \cite{Gukov:2020qaj,He:2021oav,He:2020eva,Bao:2021olg,Bao:2021ofk,Chen:2020jjw,Berman:2021mcw,Berglund:2021ztg}.
Motivated by this, we initiate the program of applying ML techniques to the classification of 5-brane webs and 5d SCFTs, concentrating on the simplest case of webs with exactly three external legs. The goal will be to teach a Siamese Neural Network (SNN) \cite{bromley:1993}, an architecture designed to determine the similarity of inputs, to recognize webs equivalent under both $SL(2,\mathbb{Z})$ and HW moves (the same architecture was proposed to condense the string landscape in \cite{He:2021eiu}). 

In order to teach the computer the (in)equivalence of webs, we constructed a controlled dataset by implementing by hand $SL(2,\mathbb{Z})$ and HW moves on webs known to be inequivalent --as their classifiers following \cite{DeWolfe:1998eu} are different. This problem turns out to be too complicated for the computer, which performs as well as random guessing. We obtained much better results by loosening the notion of equivalence. Indeed, by declaring that webs are equivalent if they share the same classifiers --which is a necessary but not sufficient condition for true equivalence-- we constructed a dataset used to train and test the SSN on which it performed far better. We stress that the SSN did not explicitly see the classifiers used to construct the dataset. More explicitly, the SSN is only fed raw web data together with the information of equivalence/non-equivalence under the loosened version for training. From that, the web learns from the examples to identify equivalent/non-equivalent pairs in the independent test data under the loosened equivalence relation, and given that we know the ultimate existence of the classifiers discriminating webs --which we actually used to construct the dataset in the first place-- it is natural to conclude that the SSN is somehow reconstructing the existence of the classifiers identified in \cite{DeWolfe:1998eu}. This is an encouraging conclusion, also supported by topological data analysis, where a degree of clustering in the embeddings of the webs produced by the SSN is observed for the case of the loosened equivalence relation, as shown in Appendix \ref{sec:tda}. In turn, this is to be compared to the failure of the SSN when faced with the problem of identifying equivalent/non-equivalent webs under the real equivalence relation. This lack of success can be turned around and used to infer an interesting lesson about the physics of sets of 7-branes, as it suggests that the full set of classifiers discriminating the true equivalence classes --if existent at all-- are much more subtle. This work therefore serves as a prime example of how, in addition to helping us perform difficult computations, \textit{ML also has the power to detect new patterns in mathematical data and show us that our current theories may be incomplete}.  

This work initiates the application of ML to the study of branewebs and 5d SCFT's, and it leaves the door open for future development. In particular, it would be very interesting to improve the performance of ML on the full problem using the exact notion of equivalence --rather than the loosened version in terms of equal invariants. This may be achieved through consideration of more complicated architectures, such as graph neural networks or generative adversarial networks. One could also look for more suitable invariants of equivalence using unsupervised learning methods. Furthermore, since this work only considers classifying brane webs with three external legs, which is equivalent to classifying sets of 7-branes, as obvious next step would be to extend this to look at brane webs with more than three legs. Moreover, this work only scratches the surface of applications, as ML may provide new techniques to tackle very interesting problems. Besides the classification problem itself for webs with arbitrarily many external legs, it may be possible to apply ML to the characterization of the different phases of a given theory --such as \cite{Bergman:2020myx}-- or the construction of the corresponding magnetic quivers (see \textit{e.g.} \cite{Cabrera:2018jxt,vanBeest:2020kou}).

%%%%%%%%%%%%%%%%%%%%%

\section{Brane Web Data}\label{sec:webdata}

Brane webs are configurations of linked $(p,q)$ 5-branes in type IIB String Theory \cite{Aharony:1997ju,Aharony:1997bh}. These branes are objects with $p$ units of magnetic charge under the RR 2-form and $q$ units of magnetic charge under the NSNS 2-form of Type IIB String Theory. Supersymmetry requires that all the 5-branes share the $x^0,\dots,x^4$ directions. Then, D5-branes (which correspond to (1,0) 5-branes) must be extended in the $x^5$ direction and pointlike in $x^6,\dots,x^9$; NS5-branes (which correspond to (0,1) 5-branes) must be extended along $x^6$ and pointlike in $x^5,x^7,\dots,x^9$; and general $(p,q)$ 5-branes must be extended along a line in the $(x^5,x^6)$ plane with slope $q/p$ and pointlike in the remaining directions. Moreover, different types of 5-branes can meet and recombine as long as charge conservation is preserved at the junction. Therefore, the resulting system forms a web in the $(x^5,x^6)$ plane. We will be working in the convention where all the charges are incoming at a junction; therefore the condition will be $\sum_{\text{junction}}p_i=\sum_{\text{junction}}q_i=0$.

We assume that the brane webs end on 7-branes \cite{DeWolfe:1999hj}. Indeed, 7-branes can be added without breaking supersymmetry (with some caveats to be discussed below) provided they are pointlike in the plane of the web and extended in the other directions. Since $(p,q)$ 5-branes can end on $[p,q]$ 7-branes, we can always assume our webs with $L$ external legs to end each on the appropriate $L$ external 7-branes.\footnote{When labelling 5-,7-branes we assume $p,q$ to be coprime. Otherwise, $(p,q)={\rm gcd}(p,q)\,(p',q')$, which represents ${\rm gcd}(p,q)$ 5-branes of type $(p',q')=(p/{\rm gcd}(p,q),q/{\rm gcd}(p,q))$  --and the same for 7-branes. The number of branes is often called \textit{multiplicity}.} The positions of the 7-branes naturally correspond to the mass-deformations of the 5d SCFT. As the location of a 7-brane along the external leg is not a parameter of the low energy theory, each 7-brane provides in principle one real parameter associated to its position. However, charge conservation and the symmetries of the plane of the web fix 3 such parameters. Thus, for a web with $L$ external legs --\textit{i.e.}, $L$ external 7-branes-- there are $L-3$ real mass deformations. A very important feature of 7-branes is that they act as magnetic sources for the string theory axiodilaton, which will be proportional to the complex logarithm $\log(x^5+i x^6)$. The branch cut is understood as a monodromy transformation: when crossing it fields undergo the appropriate action of type IIB $SL(2,\ZZ)$ duality.  %to which we will come back below.
Since the position of the cut is immaterial, one can always choose to orient the monodromy cut of all 7-branes so that it goes away from the brane web, such that none of the 5-branes cross it. 

The presence of 7-branes places extra constraints on the web in order to preserve supersymmetry. Introducing the \textit{self-intersection} $\mathcal{I}$ defined as (we only quote the version with exclusively external 7-branes)
\begin{align}\label{eq:I}
    \mathcal{I}= \left|\sum_{1\leq i<j\leq L} \det\left(\begin{array}{cc}
        p_i & p_j \\
        q_i & q_j
    \end{array}\right) \right| - \sum_{i=1}^{L}\left[\gcd(p_i,q_i)\right]^2\,,
\end{align}
the condition for supersymmetry is \cite{Iqbal:1998xb,Bergman:2020myx}
\begin{equation}
\label{eq:SUSY_condition}
    \mathcal{I}\ge-2\,.
\end{equation}
Since the dimension of the Coulomb branch --\textit{i.e.} the rank-- of the 5d SCFT is related to $\mathcal{I}$ as
\begin{align}\label{eq:rank}
    d_{\rm CB} = \frac{\mathcal{I}+2}{2}\,,
\end{align}
the supersymmetry condition can be recast in a more physical way as demanding the dimension of the Coulomb branch to be an integer greater than or equal to zero. 

Given these ingredients, our goal is to find a classification of 5-brane webs. To establish some order, we imagine fixing the number of external legs $L$ and constructing all webs with $L$ legs. In 5d SCFT language, this corresponds to constructing all possible 5d SCFTs which admit a fixed number of $L-3$ mass-deformations. Of course, there will be infinitely many such webs, since for a given $L$, there can be theories of arbitrary rank. However, even after fixing $d_{\rm CB}$, there will be infinitely many possible webs consistent with the specified data. The reason is that, as mentioned in the main text, the space of webs is hugely redundant\footnote{Of course, there is a more mundane equivalence under relabelling of the legs. % that is, under exchanging of the columns in \eqref{eq:web_matrix}.
While it is not difficult to take care of this problem, we shall also use ML tools to account for it.}:

\begin{enumerate}

\item \textbf{S-duality of type IIB string theory:} We can act on the 5-branes and 7-branes of our setup with an $SL(2,\ZZ)$ transformation. This amounts to changing all the $p$ and $q$ charges by
\begin{align}
    \left(\begin{array}{c}
         p_i \\
         q_i 
    \end{array}\right)\,\mapsto\,
    \left(\begin{array}{c}
         p_i' \\
         q_i' 
    \end{array}\right)=      \left(\begin{array}{cc}
         a & b\\
         c & d
    \end{array}\right)  \left(\begin{array}{c}
         p_i \\
         q_i 
    \end{array}\right)\,,
\end{align}
with $a,b,c,d\in\ZZ$ and $ad-cb=1$. Note that the charges of all the 7-branes are acted upon by the same $SL(2,\ZZ)$ matrix at the same time (\textit{i.e.} $\forall \, i=1\dots, L$).

\item \textbf{Hanany-Witten moves:} As mentioned, the position of the 7-brane along the external 5-brane has no effect on the low energy theory. This includes the possibility of moving the 7-brane to the other side of the 5-brane junction. If this is the case, two things happen: first, when the 7-brane crosses the junction, several 5-branes are created or annihilated; this is the well known Hanany-Witten transition. Second, in order to arrive back at the situation where the monodromy cut of the 7-brane goes away from the web without meeting any of the other objects, we need to sweep it 180 degrees. In doing this, the monodromy will act on one of the legs which we have not moved via an $SL(2,\ZZ)$ transformation (see Figure \ref{fig:brane_webs_and_HW_moves}). The action of the monodromy of a $[p,q]$ 7-brane, if we sweep it clockwise, is given by
\begin{align}
    M_{(p,q)}=\left(\begin{array}{cc}
       1-pq  & p^2 \\
        -q^2 & 1+pq
    \end{array}\right)\,,
\end{align}
and by $M_{(p,q)}^{-1}$ if we sweep it anti-clockwise. Note that in this case only one of the other external legs of the web is undergoing an $SL(2,\ZZ)$ transformation.

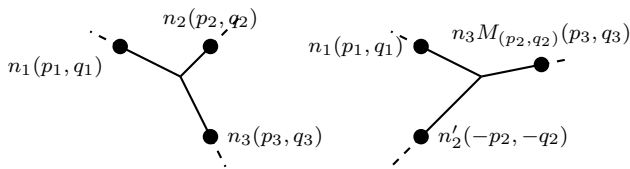
\begin{figure}[t]
    \begin{tikzpicture}[scale=0.4]
    \draw[thick](0,0)--(2,-1);
    \draw[thick](2,-1)--(3,0);
    \draw[thick](2,-1)--(3,-3);
    \draw[thick,dashed](0,0)--(-1,0.5);
    \draw[thick,dashed](3,0)--(4,1);
    \draw[thick,dashed](3,-3)--(3.5,-4);
    \node[label={[label distance=.01cm]190:{\footnotesize $n_1 (p_1,q_1)$}}][7brane]at(0,0){};
    \node[label={[label distance=.01cm]90:{\footnotesize $n_2 (p_2,q_2)$}}][7brane]at(3,0){};
    \node[label={[label distance=.01cm]0:{\footnotesize $n_3 (p_3,q_3)$}}][7brane]at(3,-3){};
    \draw[thick](10,0)--(12,-1);
    \draw[thick,dashed](10,0)--(9,0.5);
    \draw[thick](12,-1)--(10,-3);
    \draw[thick,dashed](10,-3)--(9,-4);
    \draw[thick](12,-1)--(14,-0.6);
    \draw[thick,dashed](14,-0.6)--(15,-0.4);
    \node[label={[label distance=.01cm]180:{\footnotesize $n_1 (p_1,q_1)$}}][7brane]at(10,0){};
    \node[label={[label distance=.01cm]0:{\footnotesize $n_2' (-p_2,-q_2)$}}][7brane]at(10,-3){};
    \node[label={[label distance=.01cm]90:{\footnotesize $n_3 M_{(p_2,q_2)}(p_3,q_3)$}}][7brane]at(14,-0.6){};
     \end{tikzpicture}
    \caption{Two brane webs equivalent by a HW move. The 5-branes are depicted as the lines in the $(x^5,x^6)$ plane; the dots represent the 7-branes; and the dashed lines the corresponding monodromy cut. In going from the web on the left to the one on the right, we have moved the 7-brane with label $i=2$ to the bottom of the junction. Sweeping with the monodromy clockwise changes the charges of the $[p_3,q_3]$ 7-brane by the action of $M_{(p_2,q_2)}$. After the transition, the new number of 5-branes $n_2'$ that hang from the $[p_2,q_2]$ 7-brane is determined by charge conservation. Note that we have changed the signs of the charges of the second 7-brane so that all the charges are ingoing.}
        \label{fig:brane_webs_and_HW_moves}
\end{figure}

\end{enumerate}

Thus, in order to achieve our goal of classifying 5-brane webs we need to further mod out the possible webs with fixed $L$ by $SL(2,\mathbb{Z})$ and HW moves. However, while if given a web it is easy to construct many equivalent webs, the reverse problem of finding whether or not two given webs are equivalent is in general very difficult. 

For the purposes of this work, the data specifying a brane web can be encoded in a \emph{web matrix}, which lists the charges of the external $[p_i,q_i]$ seven-branes, as well as the number $n_i$ of five-branes hanging from each of them.
\begin{align}\label{eq:web_matrix_general}
    W=\left(\begin{array}{cccc}
        n_1 p_1 & n_2 p_2 & \cdots & n_L p_L \\
        n_1 q_1 & n_2 q_2 & \cdots & n_L q_L
    \end{array}\right)\,,
\end{align}
where $L$ is the number of external legs of the web, and we work with the convention that all the charges of the five-branes are ingoing and ordered anticlockwise around the junction.

At this point it is very interesting to note that the problem of classifying branewebs contains as a sub-problem that of classifying sets of 7-branes, which has a long history in the String Theory literature (and has ramifications to the mathematics of $SL(2,\mathbb{Z})$). Indeed, given that a necessary condition (though not sufficient, as more than one web can in principle be hung from the same 7-brane set) for two webs to be equivalent is that their respective sets of external 7-branes are equivalent; as a partial step, we could consider the external 7-branes alone, which are actually subject to the same $SL(2,\mathbb{Z})$ and HW ambiguities. This problem was studied in \cite{DeWolfe:1998eu}, where it was conjectured that inequivalent sets of $L$ 7-branes are characterized by the \emph{total monodromy} of the web, defined as the product of the monodromies of all the 7-branes,
\begin{align}\label{eq:M_tot}
    M_{tot}=M_{(p_1,q_1)}M_{(p_2,q_2)}M_{(p_3,q_3)}\,,
\end{align}
and the \emph{asymptotic charge invariant} $\ell$, defined as
\begin{align}\label{eq:l}
    \ell=\gcd\left\lbrace \det\left(\begin{array}{cc}
        p_i & p_j \\
        q_i & q_j
    \end{array}\right)\,,\,\forall\, i,j \right\rbrace\,.
\end{align}

There are cases, however, where webs with the same classifiers give rise to different low energy theories. For instance, take the webs following 3-leg webs
\begin{align}
 & W_1=\left(\begin{array}{ccc}
        -4 & 3   & 1 \\
        18 & -18 & 0
\end{array}\right)\,,\nonumber\\
 & W_2=\left(\begin{array}{ccc}
        3 & -4   & 1 \\
        18 & -18 & 0
\end{array}\right)\,,\nonumber
\end{align}
corresponding to rank 2 theories. Even though these webs have the same classifiers -- that is, the same number of external 7-branes, the same $\ell$ and the same total monodromy up to $SL(2,\mathbb{Z})$ --, there is no combination of $SL(2,\mathbb{Z})$/HW moves transforming one web into the other. In this particular case, $W_{1,2}$ are related by inversion of the y-axis in the plane of the web corresponding to parity in the 5d SCFT, therefore they are physically equivalent, yet by our definition the webs are not strongly equivalent. We choose to define strong equivalence without parity because this then also tells us about the classification of sets of 7-branes, in addition to 5d SCFTs. However, we believe that this phenomenon of brane webs being inequivalent despite having equal classifiers is generic. For example, consider the following 4-leg webs:
\begin{align}
  &  W_{(c)}=\left(\begin{array}{cccc}
        -3 & 0  & 2 & 1 \\
         1 & -1 & -1 & 1
    \end{array}\right)\,, \nonumber \\
 & W_{(e)} = \left(\begin{array}{cccc}
        -3 & 1  & 1 & 1 \\
         1 & -2 & 0 & 1
    \end{array}\right)\,, \nonumber
\end{align}
belonging to different theories (c) and e) in Figure 3 in \cite{Saxena:2020ltf}). Although it was already known that these webs correspond to different theories, the observation that they share the same classifiers is new and therefore stands as a valid counter example to the classification of equivalent brane webs via invariants.  Thus, generically, the set of classifiers signaled in \cite{DeWolfe:1998eu} form a necessary condition for two 7-branes to be equivalent (instead of a sufficient condition).

Since this paper serves as a proof of concept of the applicability of ML techniques to the classification of branewebs and 5d SCFTs, we restrict to the simplest case of $L=3$. In this case charge conservation is enough to guarantee that, given a set of three 7-branes, only a unique web can be hung --up to the trivial possibility of considering multiple copies of it. Thus, for three-leg webs, their classification coincides as well with that of sets of three 7-branes\footnote{Restricting to webs with three legs has another technical advantage. Strictly speaking, the supersymmetry condition \eqref{eq:SUSY_condition} is only valid for irreducible junctions, which are those which allow realisation of the full space of $L-3$ \textit{a priori} possible mass-deformations \cite{Bergman:2020myx}. This represents an issue (whose nuances are currently not fully understood) which is not present for webs with $L=3$.}. 

%%%%%%%%%%

With this in mind, we define the following two notions of equivalent webs:
\begin{itemize}
    \item \textbf{Strong equivalence:} Two webs are strongly equivalent if they can be transformed into each other by means of any combination of $SL(2,\ZZ)$ and HW moves.
    
    \item \textbf{Weak equivalence:} Two webs are weakly equivalent if they have the same invariants \eqref{eq:I}, \eqref{eq:M_tot} (mod $SL(2,\ZZ)$) and \eqref{eq:l}.
\end{itemize}
Of course, if two webs are strongly equivalent they are also weakly equivalent.

Our goal is to teach a Neural Network to distinguish webs according to these two notions, by means of exposing it to big sets of webs where such equivalence is known. To begin with, note that webs with three external legs are described by 9 variables: $p_1,p_2,p_3,q_1,q_2,q_3,n_1,n_2,n_3$, that define the web matrix $W$ \eqref{eq:web_matrix_general}. These variables must satisfy the following three conditions: (1) $(p_i,q_i)$ must be coprime; (2) $\sum{p_i} = \sum{q_i} = 0$; and (3) the self-intersection \eqref{eq:I} satisfies $\mathcal{I} \geq -2$. We then define the following datasets: 

\begin{itemize}
    \item $\mathbf{X}$: we generate all possible $W$, for $p_i,q_i \in [-3,3]$ and $n_i \in [1,3]$, such that (1), (2) and (3) are satisfied. For each web, $W$, we compute $M,\ell$, and rank, and using this information we sort the webs into equivalence classes, of which there are 14, under the weak equivalence defined above. We take 48 webs\footnote{We take 48 webs from each class so that each set $\textbf{X}_{I}$ is of equal size, where 48 is chosen as it is the size of the smallest equivalence class.} from each class to produce a total dataset $\textbf{X} = \cup_{I=1}^{14} \textbf{X}_{I}$ of 672 webs.\footnote{Both the weak equivalence and strong equivalence datasets contain 672 webs which is small in ML terms, however, the SNN was trained on triplets taken from these datsets and so the size of the training data was actually much larger ($>100,000$).} Members from different groups are necessarily inequivalent but members of the same group may also be inequivalent.
    \item $\mathbf{Y}$: to create our second dataset we take one member from each group $\textbf{X}_{I}$ --which are surely inequivalent-- and perform a combination of HW and $SL(2,\mathbb{Z})$ moves and shuffle the order of the columns in the web matrices to generate sets, $\textbf{Y}_I, I=1,...,14$, where again $\#(\textbf{Y}_{I})=48$, of equivalent web matrices. Combining these sets produces a total dataset $\textbf{Y}= \cup_{I=1}^{14} \textbf{Y}_{I}$ of 672 webs, where webs $W^{i},W^{j}$ are equivalent (in the strong sense as defined above) if they belong to the same class $\textbf{Y}_{I}$ and inequivalent otherwise. 
\end{itemize}

\section{Machine Learning}

Our goal is to see whether, given two webs $W^{i},W^{j}$, a Siamese Neural Network (SNN) is able to determine whether or not they are equivalent. As the name suggests, SNNs consist of two or more identical sub-networks that share the same parameters, weights, and biases. These sub-networks output feature vectors (embeddings) of the inputs that are fed to a loss function, which calculates the similarity between the inputs. Here we train an SNN that takes as input the $2 \times 3$ web matrices $W^{i}$ and produces 10-dimensional embeddings, $\textbf{x}^{i} = (x_{1}^{i},...,x_{10}^{i}) \in \mathbb{R}^{10}$ (a heuristically optimal choice), such that if two webs $W^{i},W^{j}$ are equivalent the squared Euclidean distance between their embeddings, $d(\textbf{x}^{1},\textbf{x}^{2}) = \sum_{\mu=1}^{10}{(x^{i}_{\mu}-x^{j}_{\mu})^2}$, is less than some threshold. We train our network to minimise the triplet loss \eqref{eq:tripletloss} using the Adam optimisation algorithm with triplet loss parameter $\alpha = 5$ and a learning rate of 0.0001. The triplet loss requires the input training data to be given in triplets $(A,P,N)$, where $A$ (anchor) is the reference web, $P$ (positive) an equivalent web, and $N$ (negative) a non-equivalent web. The distance between the anchor embedding and positive embedding is minimised while the distance between the anchor embedding and negative embedding is maximised. The network is trained on triplets generated from 80\% of the data over 5 epochs using a batch size of 256 and 1000 steps per epoch. Batch size, epochs and steps per epoch were chosen from running a grid search algorithm on these hyperparameters. After training, we generate embeddings for the webs in the remaining 20\% using the trained sub-network and compute the squared distance between the embeddings of all pairs in this set. If the distance is less than 1, the chosen threshold value, we determine the pair to be equivalent and inequivalent otherwise. The metrics used to measure the performance of the network are accuracy \eqref{eq:accuracy} and Matthew's Correlation Coefficient (MCC) \eqref{eq:MCC} of the pairwise equivalence predictions. We use 5-fold cross validation, whereby the network is independently trained 5 times on different 80\% partitions of the data, and tested on the remaining 20\%, such that the union of the test sets gives the full dataset. Finally, on the full dataset of embeddings we applying $k$-means clustering to cluster the webs into equivalent groups, and comparing this to the true clustering we compute the Rand Index (RI) score. For more details on the ML methods used, see Appendix \ref{sec:ML_appendix}.

The computations in this paper were carried out in \texttt{python} with the use of \texttt{Tensorflow} \cite{tensorflow} and \texttt{Scikit-Learn} \cite{scikit-learn}. The SNN was built using code from \cite{SNNgithub}. Coding scripts and data can be found at this work's corresponding GitHub: \url{https://github.com/elliheyes/MLBraneWebs.git}.

\subsection{Weak Equivalence via Invariants}
\label{sec:equivalence_from_invariants}

\begin{figure*}[!tb]
    \centering
    \begin{minipage}{\columnwidth}
        \includegraphics[scale=0.35]{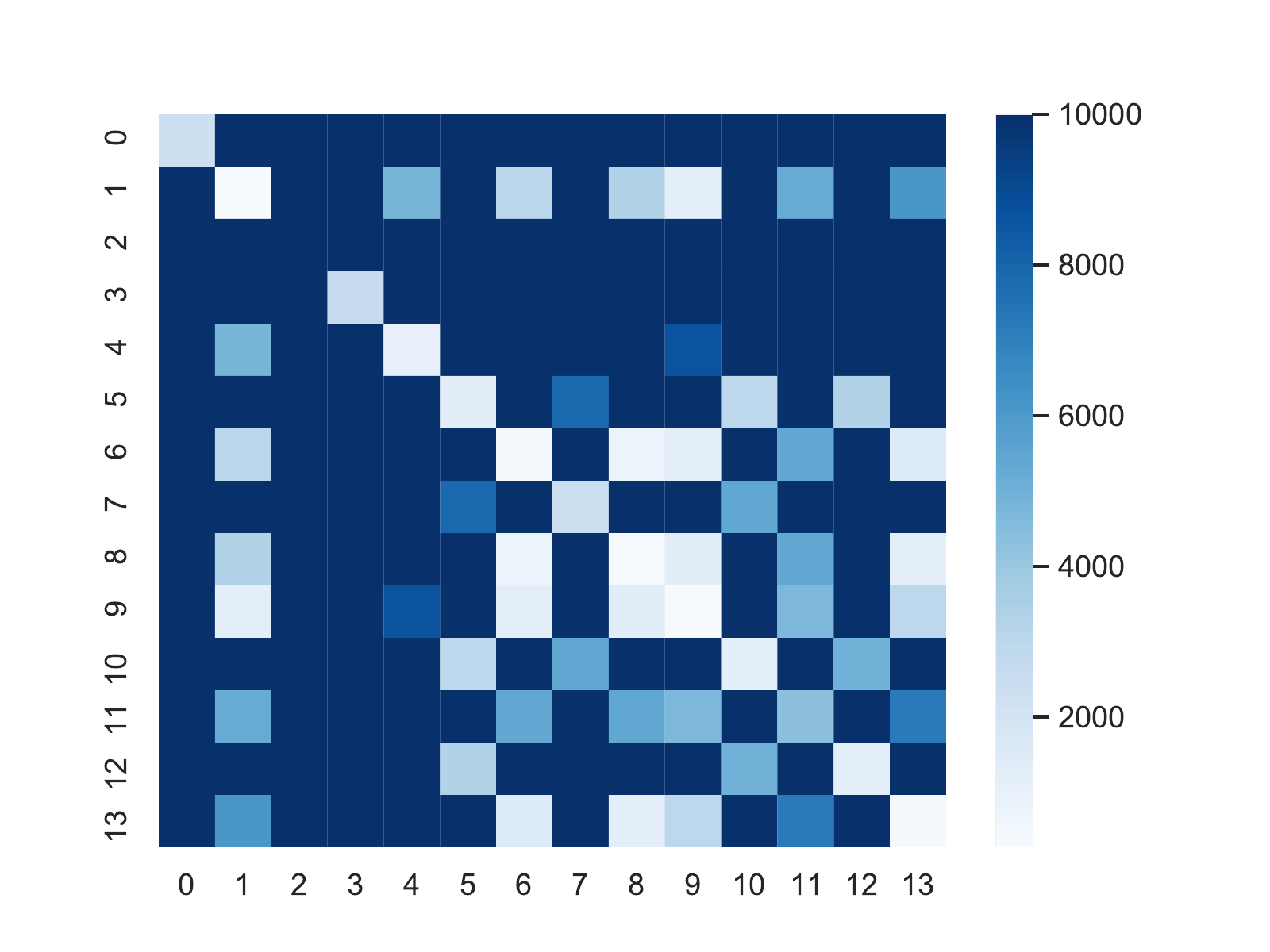}
        \caption{Mean squared Euclidean distances between webs in $\textbf{X}=\cup_{I=1}^{I=14} \textbf{X}_{I}$.}
        \label{fig:confusion_X}
    \end{minipage}
    \hfill
    \begin{minipage}{\columnwidth}
        \includegraphics[scale=0.35]{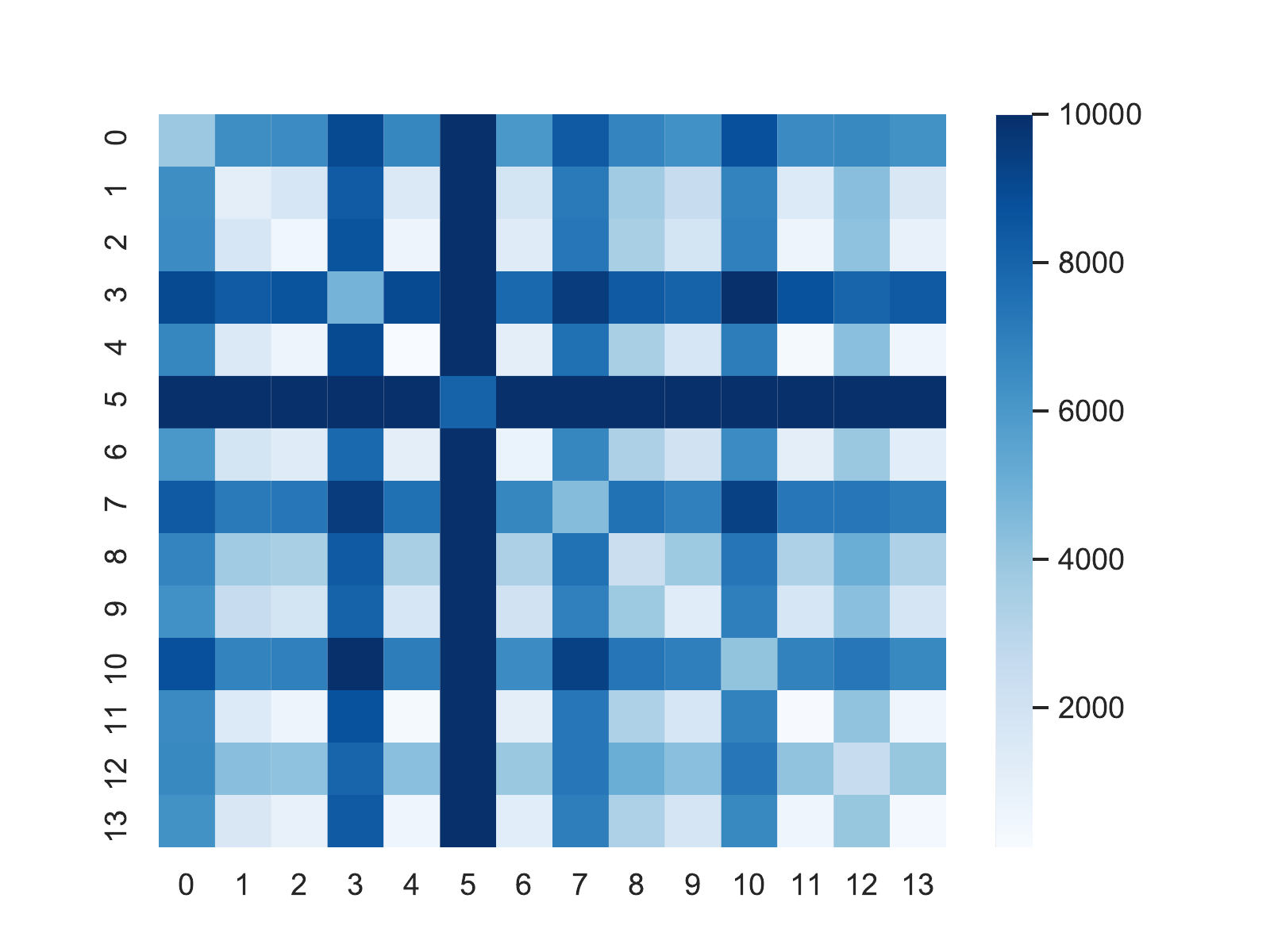}
        \caption{Mean squared Euclidean distances between webs in $\textbf{Y}=\cup_{I=1}^{I=14} \textbf{Y}_{I}$.}
        \label{fig:confusion_Y}
    \end{minipage}
\end{figure*}

We begin by considering the simplest case of just two weak equivalence groups, $\textbf{X}_{1}$ and $\textbf{X}_{2}$, and train our SNN to determine whether a pair of webs from the set $\textbf{X}_{1} \cup \textbf{X}_{2}$ are weakly equivalent, meaning they belong to the same group. We note that the choice of $\textbf{X}_{1}$ and $\textbf{X}_{2}$ is arbitrary, in that we could have chosen any two groups, but we note that the results are similar in all cases and so it suffices to use this example. The accuracy and MCC scores of the pairwise equivalence predictions made by the SNN from the web matrices in the test set are given in Table \ref{tab:SNN_results}. Amazingly we see that after training on 80\% of $\textbf{X}_{1} \cup \textbf{X}_{2}$ the network determines weak equivalence with 100\% accuracy on the remaining 20\%. The Rand Index (RI) score in Table \ref{tab:RI_results} also shows that the $k$-means clusters agree completely with the true clusters. We visualise this result in Figure \ref{fig:TSNE_X_2}, where the 10-dimensional output embeddings of the webs are reduced to 2-dimensions by t-SNE (t-distributed stochastic neighbor embedding \cite{Maaten:2008}, for an explanation see Appendix \ref{sec:ML_appendix}). The webs group together into two distinct weakly equivalent clusters. 

Motivated by this result we extend the investigation to consider the full dataset $\textbf{X}$ of 14 classes. We observe from the results in Tables \ref{tab:SNN_results} and \ref{tab:RI_results} that the performance scores have dropped to more modest values but the network is still performing significantly better than random guessing. The mean squared Euclidean distances between the embeddings of each pair of webs from the 14 classes are displayed in Figure \ref{fig:confusion_X}. We see that the distances along the diagonal (i.e. for webs belonging to the same class) are close to 0, while distances between webs from different classes are often much larger than 0, which is the desired result. This plot also helps us to identify on which classes the network performs best.

\subsection{Strong Equivalence via Generation}
\label{sec:equivalence_from_generation}

\begin{figure*}[!tb]
    \centering
    \begin{minipage}{\columnwidth}
        \includegraphics[width=0.8\linewidth]{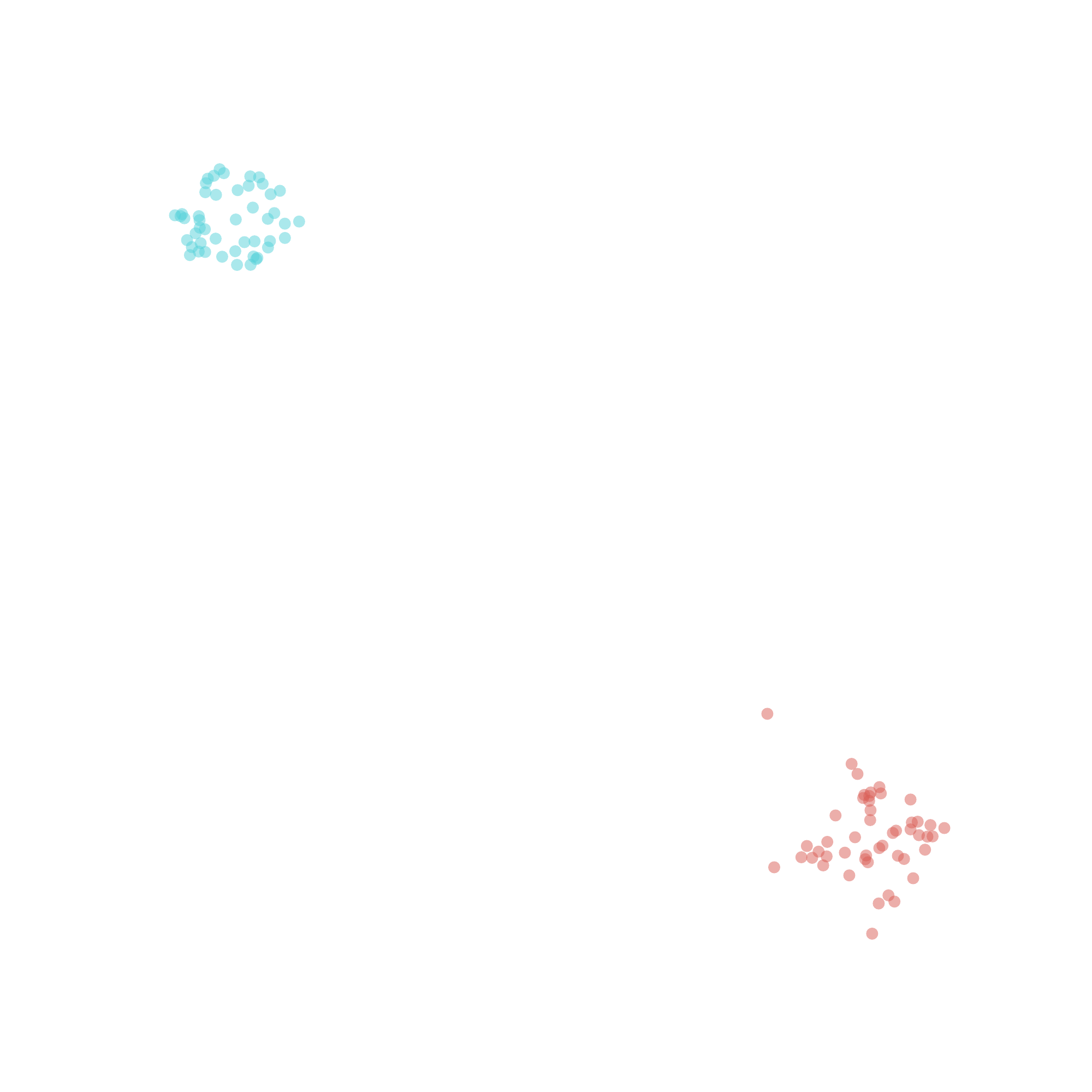}
        \caption{t-SNE $\textbf{X}_{1} \cup \textbf{X}_{2}$}
        \label{fig:TSNE_X_2}
    \end{minipage}
    \hfill
    \begin{minipage}{\columnwidth}
        \includegraphics[width=0.8\linewidth]{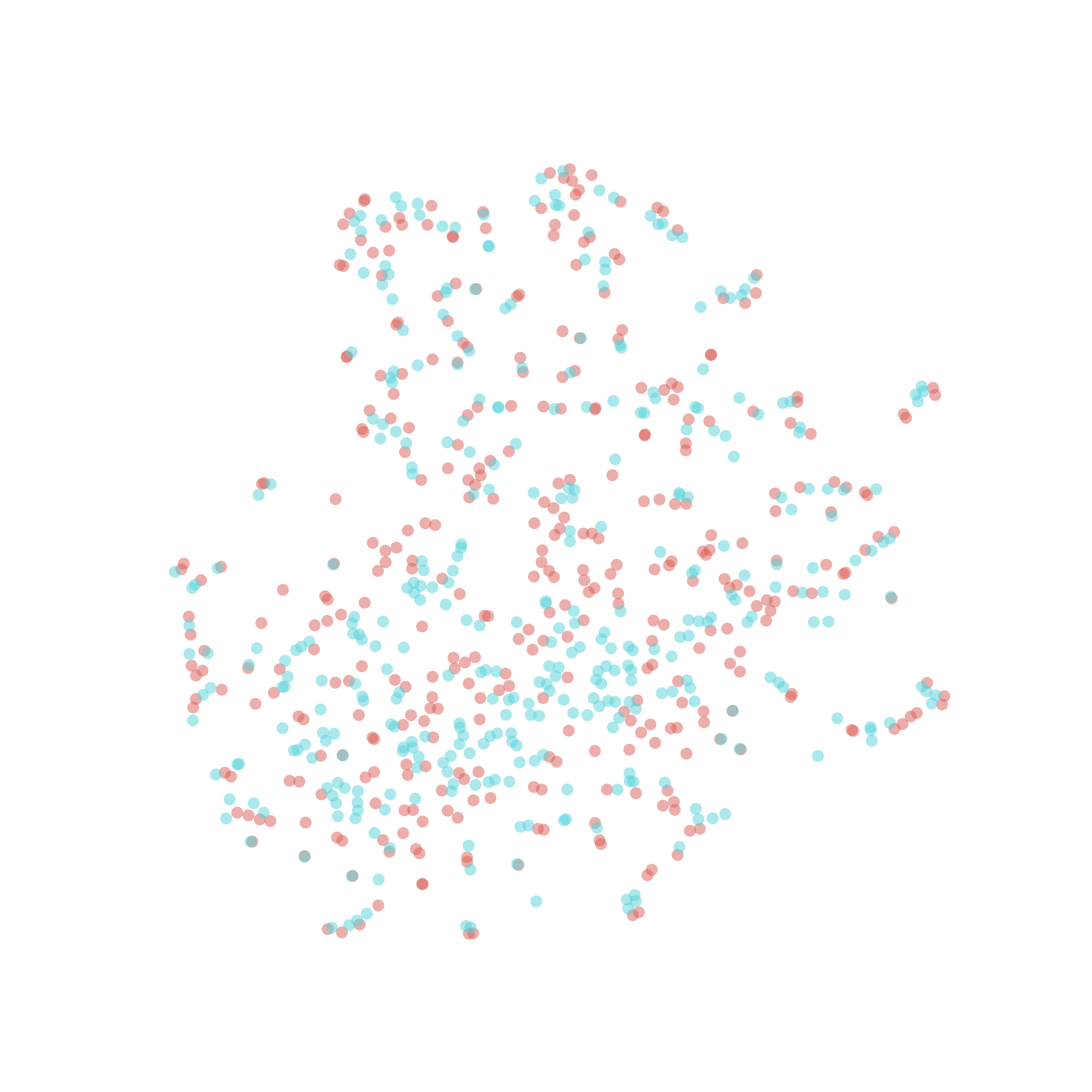}
        \caption{t-SNE $\textbf{Y}_{1} \cup \textbf{Y}_{2}$}
        \label{fig:TSNE_Y_2}
    \end{minipage} \\
    \begin{minipage}{\columnwidth}
        \includegraphics[width=0.8\linewidth]{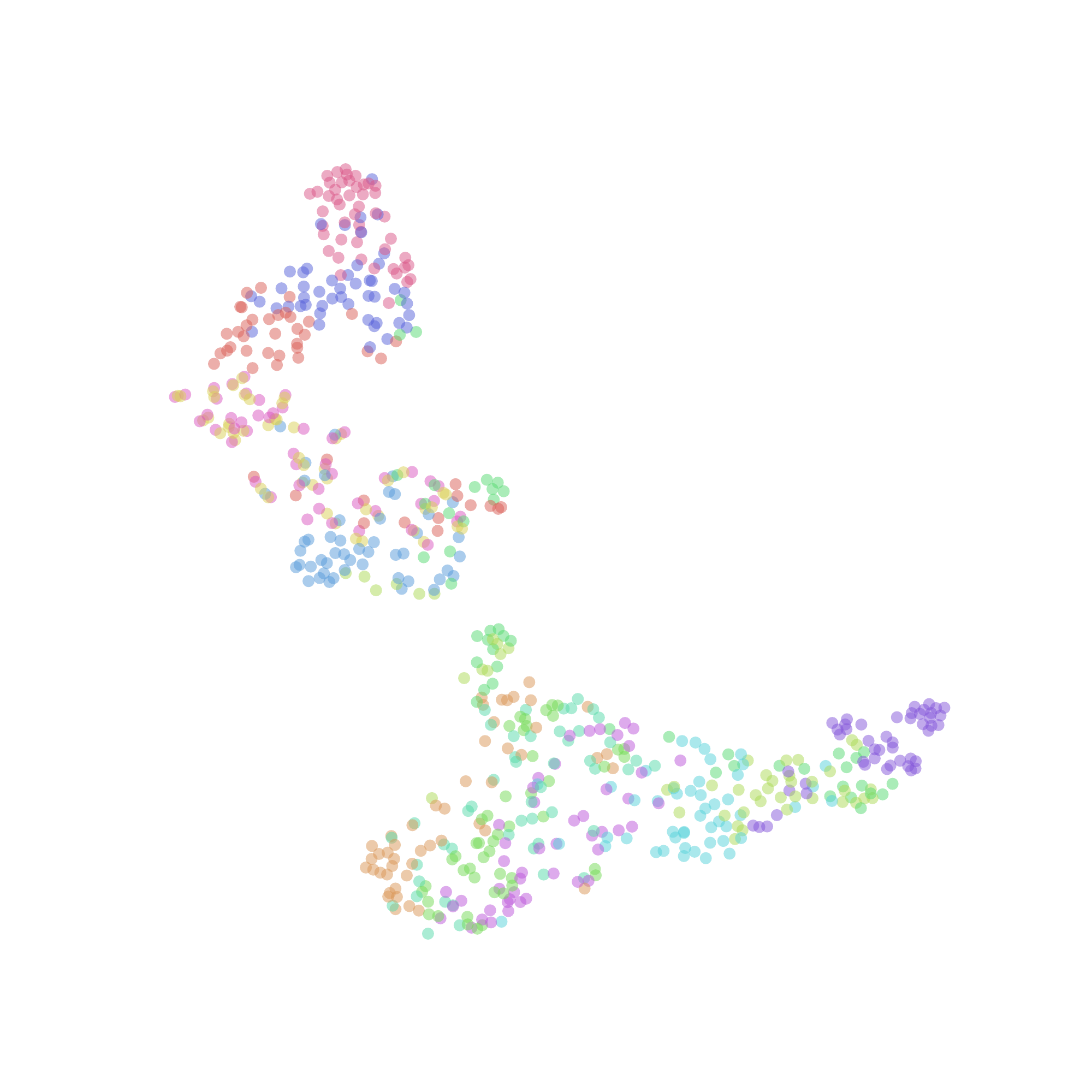}
        \caption{t-SNE $\textbf{X}$}
        \label{fig:TSNE_X_all}
    \end{minipage}
    \hfill
    \begin{minipage}{\columnwidth}
        \includegraphics[width=0.8\linewidth]{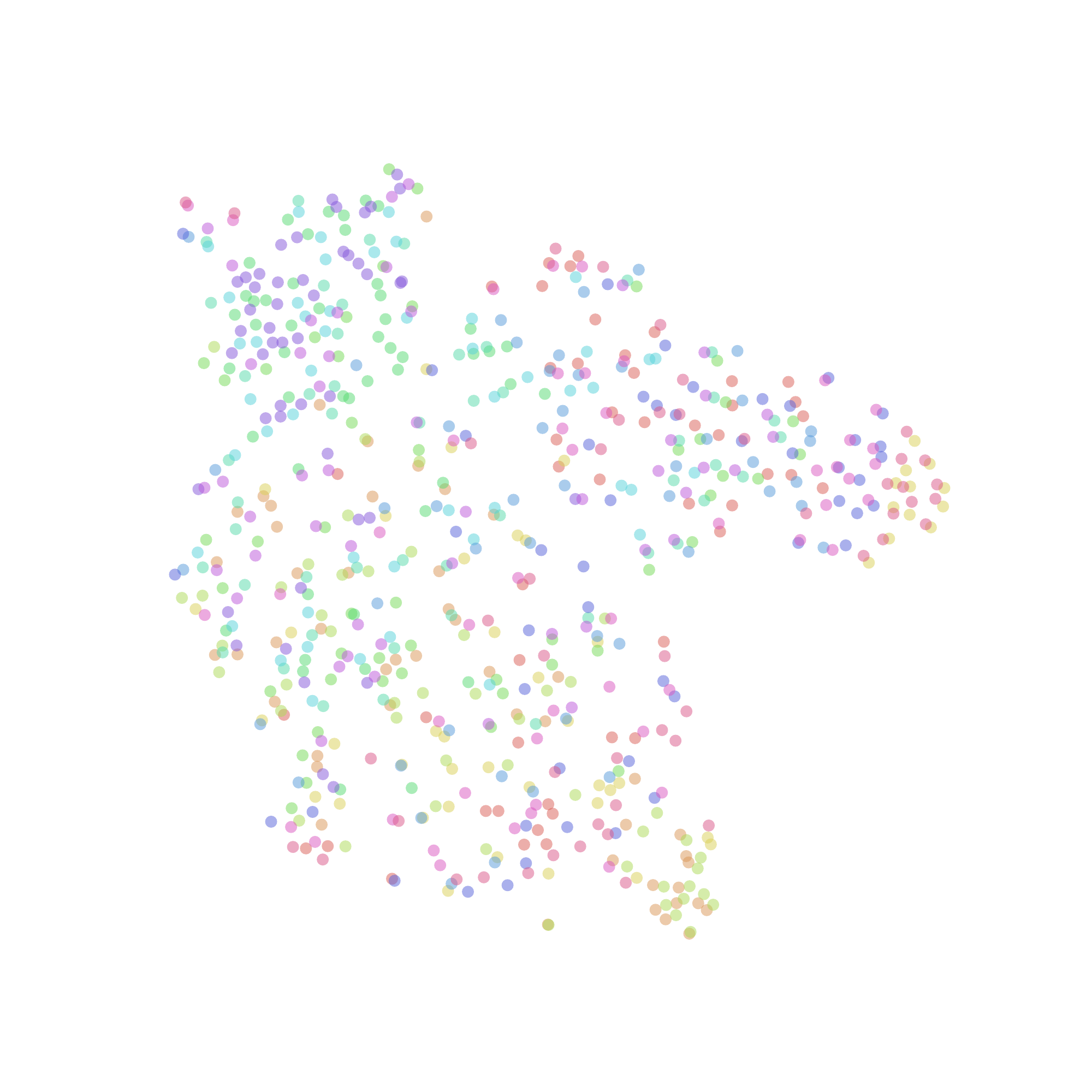}
        \caption{t-SNE $\textbf{Y}$}
        \label{fig:TSNE_Y_all}
    \end{minipage}
    \captionsetup{width=.8\linewidth}
    \caption*{t-SNE plots of the 10-dimensional web embeddings generated by the SNN of webs from the respectively labelled datasets, reduced to 2 dimensions.}
    \label{fig:t-SNE_all}
\end{figure*}

In the previous subsection we saw that an SNN is capable of identifying pairs of webs that satisfy the necessary condition for equivalence - that rank, $\ell$, and $M$, up to $SL(2,\mathbb{Z})$, are all equal. In this subsection we repeat the same procedure but consider strong equivalence, and hence the dataset \textbf{Y}. We begin as before with the simplest case of just two strong equivalence classes $\textbf{Y}_{1}$ and $\textbf{Y}_{2}$ from $\textbf{Y}$, and train the SNN to distinguish pairs of webs that belong to the same class $\textbf{Y}_{I}$. We again note that the results with any other two classes, $\textbf{Y}_{i}$ and $\textbf{Y}_{j}$, are similar. As the scores in Table \ref{tab:SNN_results} show, the network fails to give any improvement on a random guess (which held for any choice of the 2 classes to compare). The RI score in Table \ref{tab:RI_results} also supports this conclusion. It can be seen from the t-SNE plot in Figure \ref{fig:TSNE_Y_2} that the embeddings of $\textbf{Y}_{1}$ and $\textbf{Y}_{2}$ are mixed together and completely indistinguishable. Despite this poor result, we extend the investigation, as we did in Section \ref{sec:equivalence_from_invariants}, to the full dataset $\textbf{Y}$. Mean distances between the embeddings of each pair of webs are displayed in Figure \ref{fig:confusion_Y}. Comparing Figures \ref{fig:confusion_X} and \ref{fig:confusion_Y} we see that strongly equivalent webs are mapped relatively further away by the SNN than weakly equivalent webs, shown by darker elements along the diagonal, and non strongly equivalent webs are mapped relatively closer together than non weakly equivalent webs, shown by lighter elements off the diagonal. The scores in Tables \ref{tab:SNN_results} and \ref{tab:RI_results} show that the SNN is no better at determining equivalence of web pairs from $\textbf{Y}$ than $\textbf{Y}_{1} \cup \textbf{Y}_{2}$ and is still no better than random guessing. This is supported by the t-SNE plot in Figure \ref{fig:TSNE_Y_all} where we see no clusters for the classes $\textbf{Y}_{I}$.

\begin{table}[h!]
\centering
\begin{tabular}{c | c c } 
  & Accuracy & MCC \\ 
 \hline
 $\textbf{X}_{1} \cup \textbf{X}_{2}$ & $1.0000 \pm 0.0000$ & $1.0000 \pm 0.0000$ \\ 
 $\textbf{Y}_{1} \cup \textbf{Y}_{2}$ & $0.5008 \pm 0.0139$ & $0.0025 \pm 0.0324$ \\
 \textbf{X} & $0.7692 \pm 0.0997$ & $0.5523 \pm 0.2079$ \\
 \textbf{Y} & $0.4864 \pm 0.0637$ & $-0.0268 \pm 0.1305$ \\
\end{tabular}
\caption{Accuracy and MCC scores of SNN pairwise equivalence predictions on the test sets.}
\label{tab:SNN_results}
\end{table}

\begin{table}[h!]
\centering
\begin{tabular}{c | c } 
  & Rand Index \\ 
 \hline
 $\textbf{X}_{1} \cup \textbf{X}_{2}$ & $1.0000 \pm 0.0000$ \\ 
 $\textbf{Y}_{1} \cup \textbf{Y}_{2}$ & $0.5031 \pm 0.0072$ \\
 \textbf{X} & $0.8905 \pm 0.0053$ \\
 \textbf{Y} & $0.3229 \pm 0.0456$ \\
\end{tabular}
\caption{Rand index scores of $k$-means clustering of SNN web embeddings on the full datasets.}
\label{tab:RI_results}
\end{table}

\section*{Acknowledgements}
GAT is supported by the Spanish government scholarship MCIU-19-FPU18/02221.
YHH~would like to thank~STFC for grant ST/J00037X/2.
E.~Heyes would like to thank City, University of London, for a~PhD studentship and the States of Jersey for a postgraduate grant.
E.~Hirst would like to thank STFC for a~PhD studentship. DRG and GAT are partly supported by Spanish national grant MINECO-16-FPA2015- 63667-P as well as the Principado de Asturias grant FC-GRUPIN-IDI/2018/000174. DRG would like to thank O.~Bergman for conversations. 

%\FloatBarrier
\bibliographystyle{utphys}
\bibliography{references}

\appendix

\section{Topological Data Analysis}\label{sec:tda}

\paragraph*{\textbf{Raw Web Data}}
To analyse the dataset of web information, both pre and post to embedding with the SNN's base model we turn to a tool from Topological Data Analysis (TDA). 
\textit{Persistent homology} provides useful feature analysis of higher-dimensional datasets that cannot be feasibly plotted. The process plots the webs as points in their respective higher $d$-dimensional spaces, then creates a filtration of Vietoris-Rips complexes, where $\lambda$ points are connected in a $\lambda$-simplex if their $d$-dimensional balls drawn centred on each point intersect (up to $\lambda_{max}=d$). The filtration of complexes is then built during the process of the ball radii being continuously increased from 0 to $\infty$.

Throughout the filtration, $H_0$ features are all the points, born at radius 0, which then die as they become part of a connected component; \textit{i.e.} as two components combine the larger feature takes priority and survives whilst the smaller feature (usually just a point) dies.
The $H_1$ features, plotted on the same persistence diagrams, effectively represent 1d loop structures in the dataset. 
These are born as points connect to form a loop which is not a boundary of a union of 2d simplices in the complex. The feature then dies as the loop is filled in by the respective 2d simplices.
Persistence diagrams plot the features as (birth, death) pairs, and were computed with the use of the \textit{ripser} library \cite{ctralie2018ripser} in \texttt{python}. For other uses of persistent homology in theoretical physics see \cite{Cirafici:2015pky,Cole:2018emh,Berman:2021mcw}.

Each web in the raw web datasets \textbf{X} and \textbf{Y} amounts to the contents of the web matrix \ref{eq:web_matrix_general} with $L=3$. 
These 6 integers are plotted in $\mathbb{R}^6$ for the persistent homology analysis. 
Since datapoints are restricted to the integer lattice this translates to a grid-like distribution of features in the persistence diagrams, as only at specific radii can balls intersect and change the complex.

\begin{figure*}[!tb]
    \centering
    \begin{minipage}{\columnwidth}
        \includegraphics[width=0.8\linewidth]{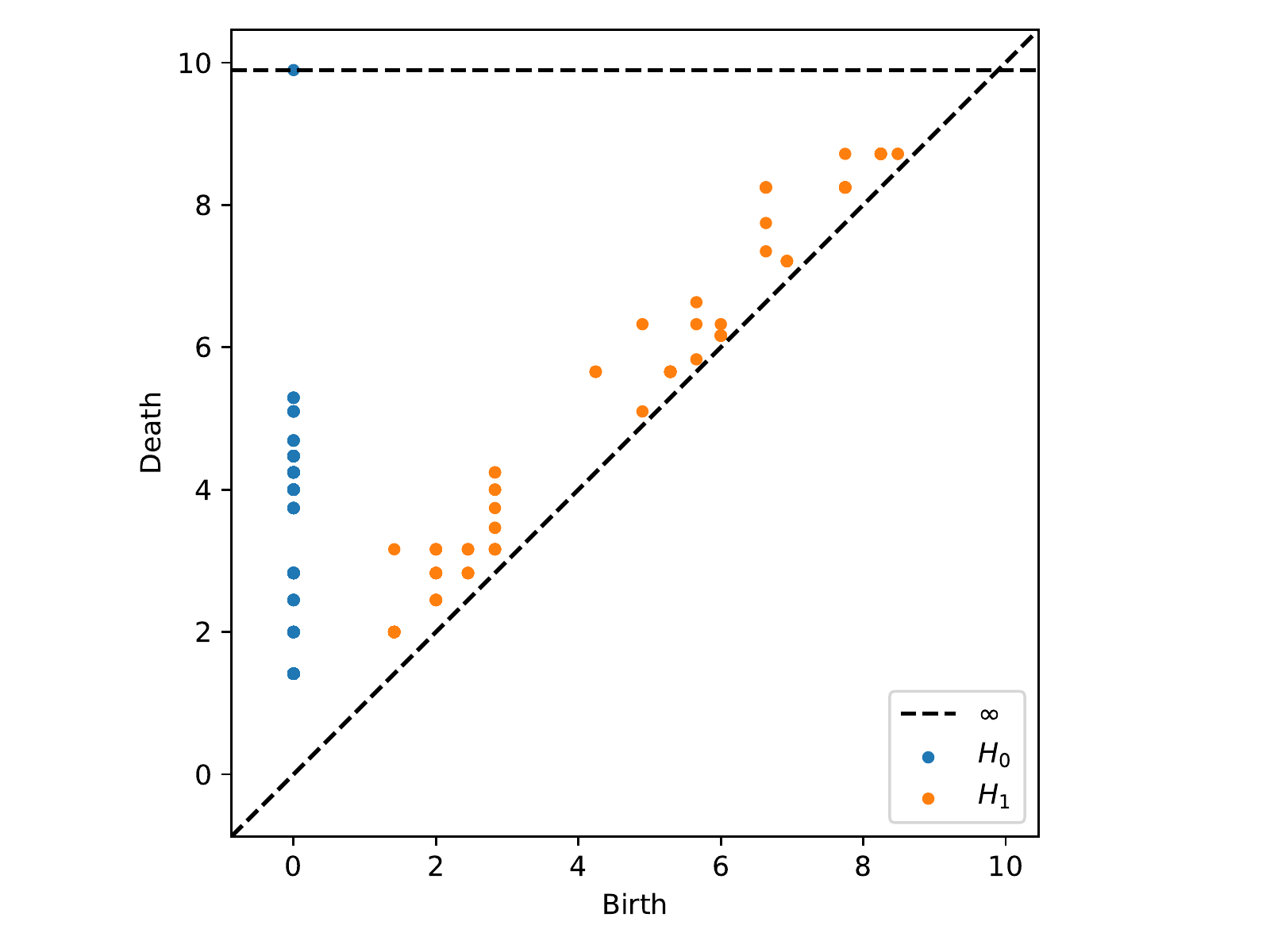}
        \caption{$(n_ip_i,n_iq_i)$ \textbf{X} data}
        \label{fig:TDA_rawX}
    \end{minipage}
    \hfill
    \begin{minipage}{\columnwidth}
        \includegraphics[width=0.8\linewidth]{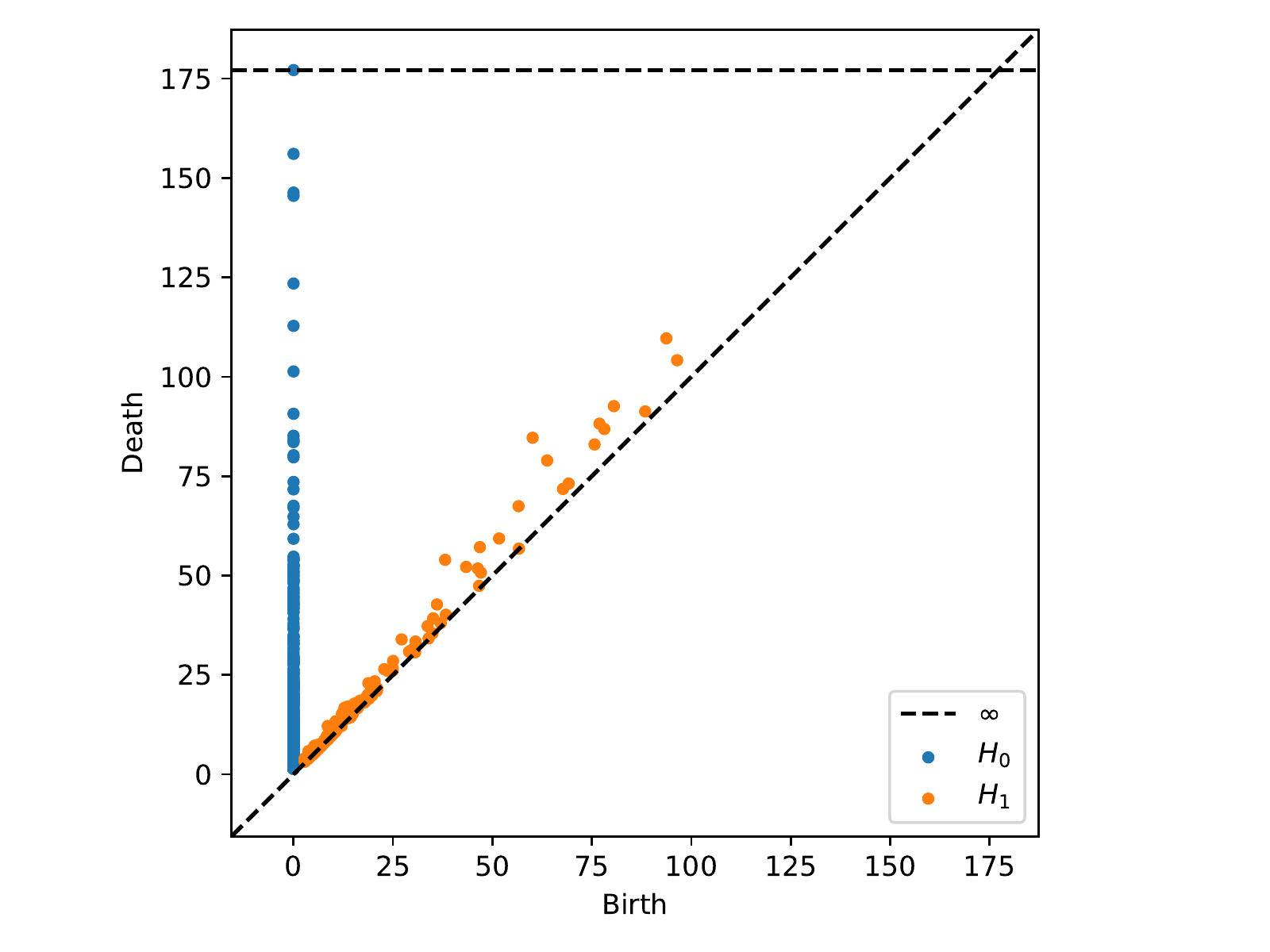}
        \caption{$(n_ip_i,n_iq_i)$ \textbf{Y} data}
        \label{fig:TDA_rawY}
    \end{minipage} \\
    \begin{minipage}{\columnwidth}
        \includegraphics[width=0.8\linewidth]{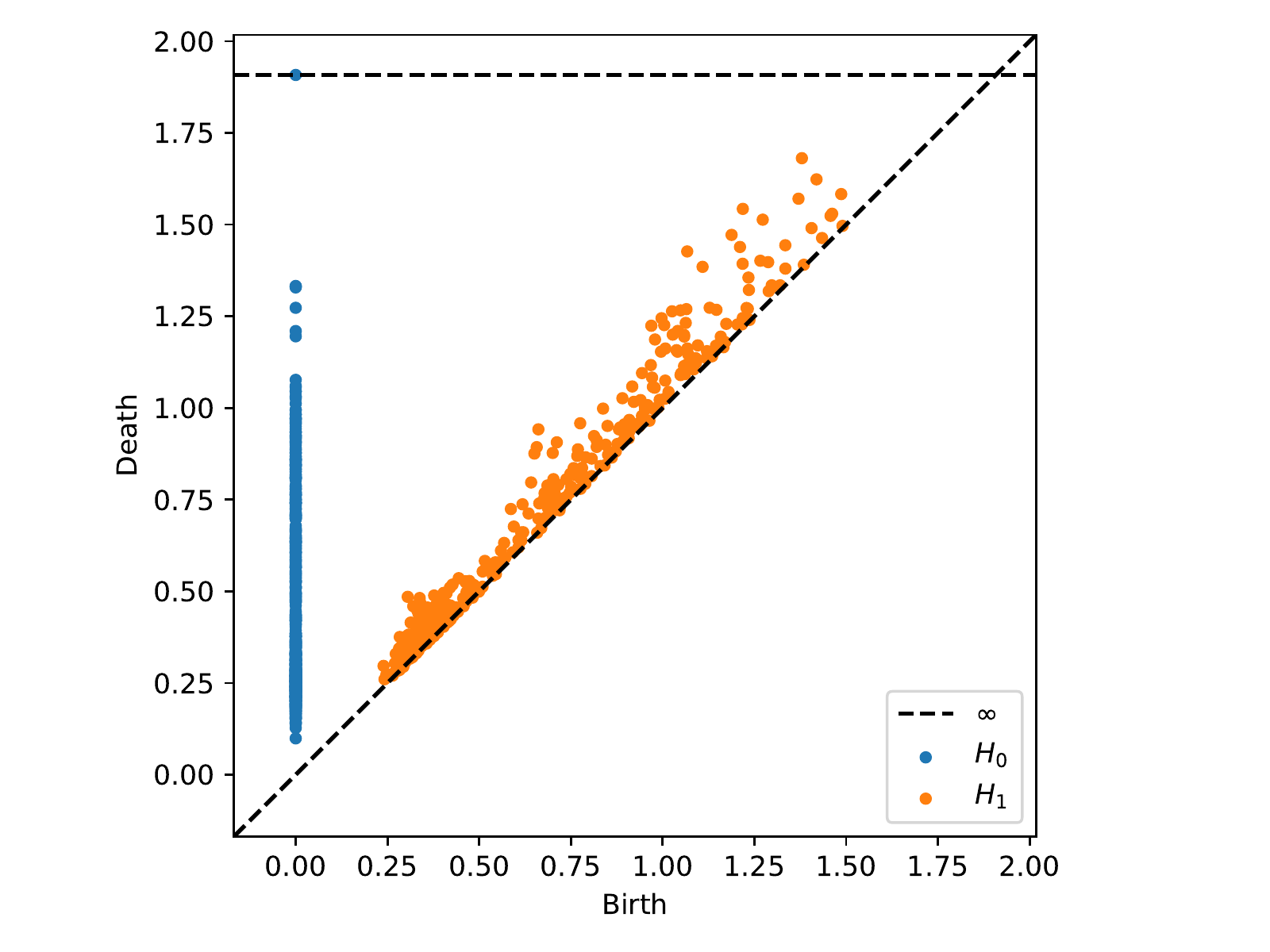}
        \caption{SNN embedded \textbf{X} data}
        \label{fig:TDA_embedX}
    \end{minipage}
    \hfill
    \begin{minipage}{\columnwidth}
        \includegraphics[width=0.8\linewidth]{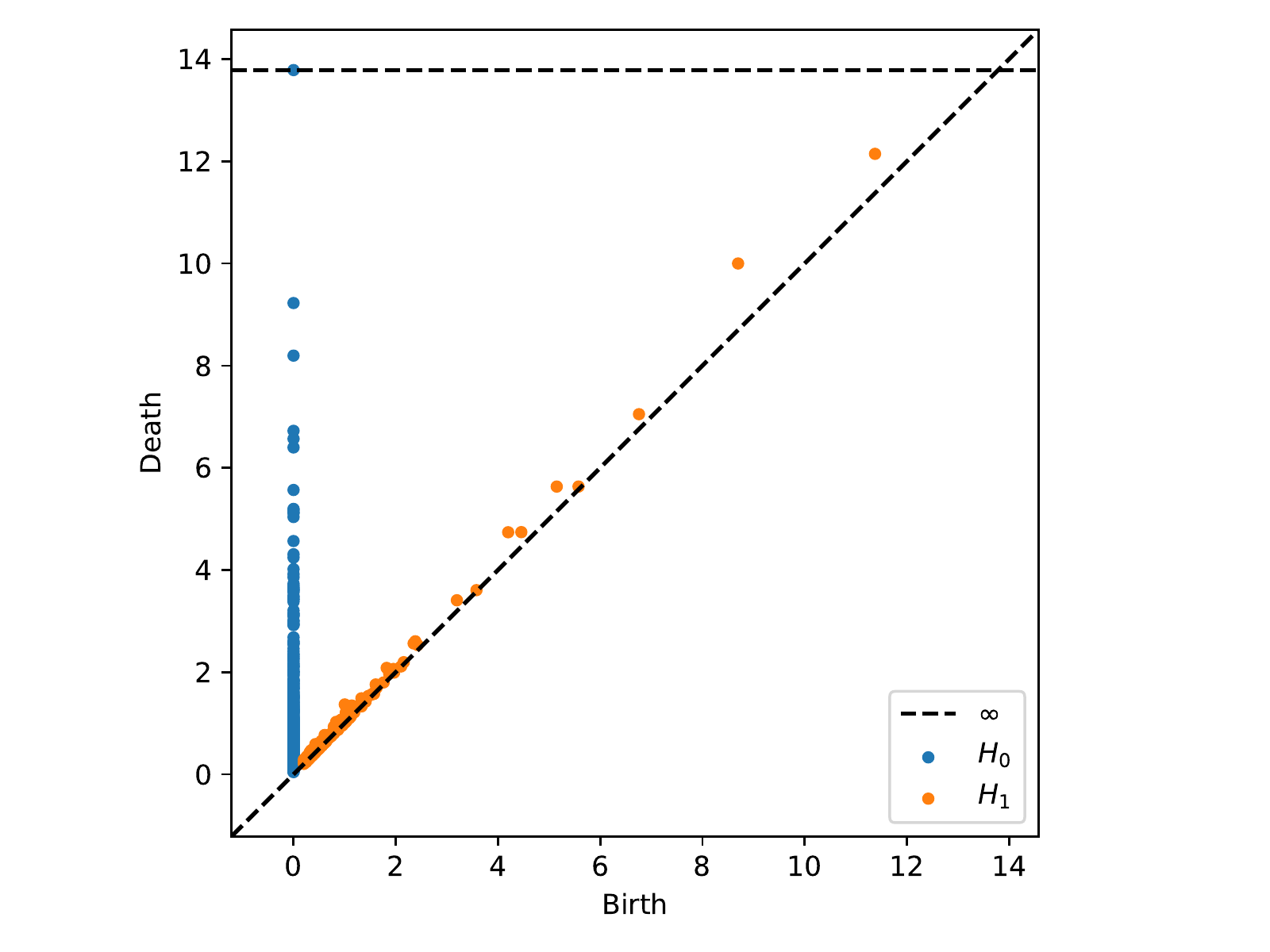}
        \caption{SNN embedded \textbf{Y} data}
        \label{fig:TDA_embedY}
    \end{minipage}
    \captionsetup{width=.8\linewidth}
    \caption*{Persistence diagrams for $H_0$ and $H_1$ on the web $(n_ip_i,n_iq_i)$ data and SNN base model embedded data, for datasets \textbf{X} and \textbf{Y} respectively.}
    \label{fig:TDA_all}
\end{figure*}

The analysis is performed for both datasets of 672 web datapoints, \textbf{X} in Figure \ref{fig:TDA_rawX}, and \textbf{Y} in Figure \ref{fig:TDA_rawY}. 
The HW moves in the generation process for \textbf{Y} lead to much larger web matrix entries, as reflected in the larger scales and finer grid structure in \ref{fig:TDA_rawY}.
For the \textbf{X} data the roughly uniform distribution of $H_0$ features indicates points are uniformly distributed in the space, as expected from the exhaustive generation procedure over the search space. Conversely the \textbf{Y} data has a slightly larger gap in the line ($\sim$125-145, note the much larger scale here) indicating there are clusters of points further from the main cluster, likely a result of some HW moves jumping datapoints away from the bulk.

Both datasets have $H_1$ features close to the diagonal, behaviour typical of noise; since there are no features far from the diagonal there is not a significant loop structure in the data which would otherwise indicate regions either omitted from the sampling or not physically plausible. \\

\paragraph*{\textbf{SNN-Embedded Web Data}}

To further analyse the success of the SNN embedding procedure, beyond the t-SNE plots, we additionally use persistent homology to examine the embedded representations of the webs.
This embedded web data is the result of the SNN's base model mapping the 6-parameter web matrix data into the $\mathbb{R}^{10}$ embedding space, where the aim of the SNN is to create an embedding model which separates inequivalent webs into their respective clusters.

The $H_0$ analysis for the embedded \textbf{X} data in Figure \ref{fig:TDA_embedX} shows a continuous stream of features indicating points merging together to form independent simplices, then the gap ($\sim$1.10-1.20) followed by a collection of features closer together indicates the separate clusters for each of the web classes combining together to finish the filtration.
Since the features are relatively close they represent more symmetrically distributed clusters, nice behaviour since a priori no equivalence classes should be especially more related than others.
This behaviour supports the success of the SNN embedding to separate the webs into clusters based on the equivalence used for training.
The $H_0$ features for the embedded \textbf{Y} data, in Figure \ref{fig:TDA_embedY}, has similar behaviour, however the cluster separation is less uniform, indicated by the less consistent line where clusters aren't combining smoothly and then only a couple of features are separated from the main line ($>8$), therefore some of the classes' clusters merge earlier, and are not as well separated as in the embedded \textbf{X} data, making this classification worse, and hence supporting the poorer learning results observed.

For both embedded datasets the $H_1$ features lie close to the diagonal, again indicating a lack of significant loop structures in the data clouds such that the clustering performs as expected.
Relatively, the embedded \textbf{X} data has many more higher birth features, likely a consequence of the better separated clusters combining later in the filtration.

\section{Machine Learning}
\label{sec:ML_appendix}

\subsection{Siamese Neural Networks}

Siamese Neural Networks (SNNs), first introduced in 1993 \cite{bromley:1993} to solve signature verification, are neural network architectures, made up of two or more identical sub-networks, that determine the similarity of inputs. The sub-networks $f_{w}$ map elements of a dataset $\mathcal{D}$ to $\mathbb{R}^{d}$, where $w$ denotes the weights and biases of the network. The goal is to train the network so that similar elements of $\mathcal{D}$ are mapped close together in $\mathbb{R}^{d}$, and dissimilar elements are mapped far apart. The $w$ are determined by extremising a loss function that is dependent on the squared Euclidean distance between the embeddings:
\begin{align}
    d_{w}(W_{1},W_{2}) \equiv (f_{w}(W_{1})-f_{w}(W_{2}))^{2}
\end{align}
where $f_{w}(W)$ denotes the embedding of input $W$. In this work we adopt the triplet loss function described below. This loss is then interpreted as our desired similarity score. 

Since we are using the triplet loss our network is made up of three identical sub-networks that produce embeddings for the anchor, positive and negative input. Each of these sub-networks consists of a 2 dimensional convolution layer, with 8 filters, a kernel size of 2, and ReLU activation functions, followed by a dense layer, with 50 neurons and ReLU activation, and the output layer is a dense layer with 10 neurons. The output of these three sub-networks is fed to an external output layer that computes the triplet loss.

\begin{figure}
    \centering
    \includegraphics[scale=0.5]{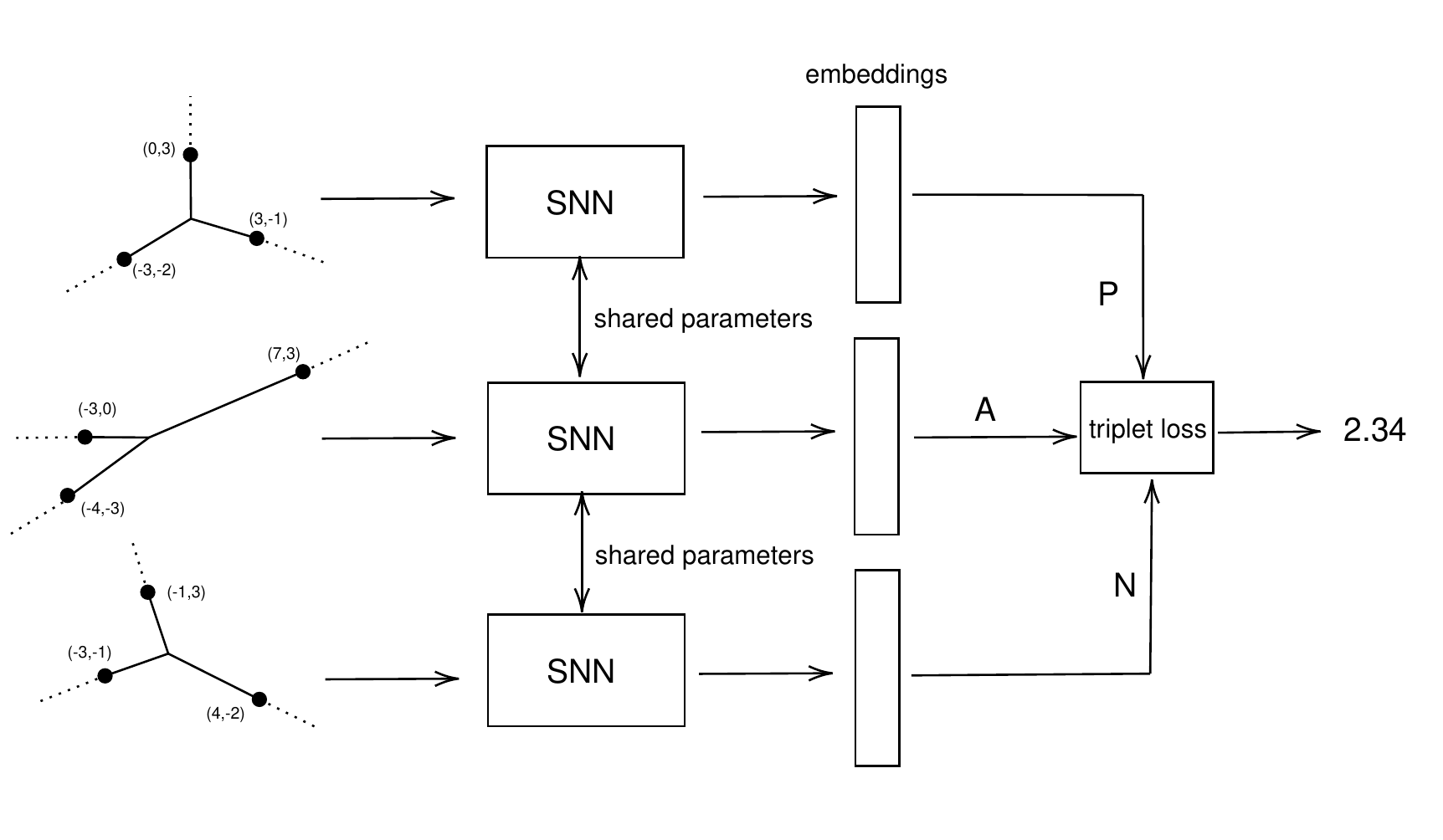}
    \caption{Diagram of Siamese Neural Netowrk architecture using triplet loss.}
    \label{fig:network_diagram}
\end{figure}

\subsection{Triplet Loss}

The triplet loss takes as input a triple: anchor (A), positive (P), and negative (N), where the anchor and positive input are similar (\textit{i.e.} equivalent webs) and the anchor and negative input are not. We want the distance between the anchor embedding and the positive embedding to be no larger than the distance between the anchor embedding and the negative embedding, that is:
\[
    ||f_{w}(A)-f_{w}(P)||^2 \leq ||f_{w}(A)-f_{w}(N)||^2 \;,
\]
Equivalently, we could write
\[
    ||f_{w}(A)-f_{w}(P)||^2 - ||f_{w}(A)-f_{w}(N)||^2 \leq 0\;.
\]
This inequality is satisfied if the $f_{w}$ always outputs 0 (or another constant value), in other words if the embeddings are all equal. To prevent the network from doing this we modify the objective so that the difference needs to be strictly less than zero by introducing a hyperparameter $\alpha \in \mathbb{R}_{>0}$:
\[
    ||f_{w}(A)-f_{w}(P)||^2 - ||f_{w}(A)-f_{w}(N)||^2 + \alpha \leq 0\;.
\]
The triplet loss function is then defined as :
\begin{equation}\label{eq:tripletloss}
\begin{split}
    \mathcal{L}(A,P,N) = {\rm max}(&||f_{w}(A)-f_{w}(P)||^2 - \\
    &||f_{w}(A)-f_{w}(N)||^2 + \alpha, 0)\;.
\end{split}
\end{equation}
This is the loss of a single triplet, the overall loss function of our network is computed as the mean of these individual losses over $\mu$ triplets:
\begin{align}
    J = \frac{\sum_{i=1}^{\mu}{L(A^{(i)},P^{(i)},N^{(i)})}}{\mu}.
\end{align}

\subsection{Evaluation Metrics}

The metrics used to evaluate the network performance, classifying equivalence of web pairs, were accuracy and Matthew's Correlation Coefficient (MCC):
\begin{align}\label{eq:accuracy}
{\rm Accuracy} = \frac{TP+TN}{TP+TN+FP+FN} \in [0,1] 
\end{align}
\begin{align}\label{eq:MCC}
\resizebox{0.5\textwidth}{!}{${\rm MCC} = \frac{TP \times TN - FP \times FN}{\sqrt{(TP+FP) \times (TP+FN) \times (TN+FP) \times (TN+FN)}} \in [-1,1]$} 
\end{align}
where $TP,TN,FP,FN$ denote the number of true positive, true negative, false positive and false negative predictions respectively. Accuracy reflects the proportion of test web pairs whose equivalence is correctly determined. The MCC is similar to the Pearson correlation coefficient, where a score of 1 indicates complete agreement between predictions and truth, 0 indicates that the predictions are no better than random guessing, and -1 indicates complete disagreement between the predicted and true equivalence. 

\subsection{K-Means Clustering}

$k$-means clustering sorts datapoints into $k$ clusters, where each datapoint belongs to the cluster with the nearest centroid. The objective of the $k$-means algorithm is to minimise the sum of squared distances between the centroid and the datapoints. Formally, given a set of datapoints $(\textbf{x}_{1},...,\textbf{x}_{n})$, $k$-means clustering aims to partition the datapoints into $k$ sets $\textbf{S}=(S_{1},...,S_{k})$ so as to minimise
\begin{align}
    \sum_{i=1}^{k}{\sum_{\textbf{x} \in S_{i}}{||\textbf{x} - \mu_{i}||^2}}
\end{align}
where $\mu_{i}$ is the centroid of the set $S_{i}$.
The first step is to choose the value for $k$, i.e. the number of clusters. Then the algorithm starts by randomly selecting $k$ points from the data to be centroids and assigning each datapoint to the cluster with the closest centroid. Then the new centroid of each cluster is computed and the datapoints are again assigned to the nearest cluster. This iterative two step process of computing the new centroid and assigning datapoints to clusters is repeated until the centroids do not change their position.

\subsection{Rand Index}

The \textit{Rand Index} is a measure of similarity between two partitions. It considers all pairs of elements, counting pairs where both elements are assigned to the same subset, or different subsets, across the partitions; used here for comparison of the predicted and true partitions. Let $X=\{x_1,...,x_n\}$ be a set of $n$ elements and $A=\{A_1,...,A_r\}$, $B=\{B_1,...,B_s\}$ be two partitions of $X$ into $r$ and $s$ subsets respectively. Then the Rand index, $R$, is given by 
\begin{align}\label{eq:RI}
    R = \frac{a+b}{a+b+c+d}
\end{align}
where 
\begin{itemize}
    \item $a$ is the number of pairs of elements in $X$ that are in the same subset in $A$ and the same subset in $B$.
    \item $b$ is the number of pairs of elements in $X$ that are in different subsets in $A$ and different subsets in $B$.
    \item $c$ is the number of pairs of elements in $X$ that are in the same subset in $A$ and different subsets in $B$.
    \item $d$ is the number of pairs of elements in $X$ that are in different subsets in $A$ and the same subset in $B$.
\end{itemize}
Therefore, $R$ exists in the range [0,1] where 1 corresponds to a perfect agreement between the partitions. 

\subsection{t-Distributed Stochastic Neighbour Embedding (t-SNE)}

t-distributed stochastic neighbour embedding (t-SNE) is a popular method, introduced in \cite{Maaten:2008}, to reduce high-dimensional data by embedding it in a low-dimensional space. It calculates similarities between datapoints in the high dimensional space and in the low dimensional space and tries to minimise the divergence between the two. 

We can break t-SNE down into three steps:
\begin{enumerate}
    \item Step 1: Given a set of $N$ points, $\textbf{x}_{1},...,\textbf{x}_{N}$ in high dimensional space, measure the similarities between points $\textbf{x}_{i}$. The similarity of datapoint $\textbf{x}_{i}$ to datapoint $\textbf{x}_{j}$ is the conditional probability, $p_{j|i}$, that $\textbf{x}_{i}$ would pick $\textbf{x}_{j}$ as its neighbour. This probability is proportional to the probability density under a Gaussian centred at $x_{i}$.
    \begin{align}
    p_{j|i} = 
        \begin{cases}
            \frac{\text{exp}(-||\textbf{x}_{i} - \textbf{x}_{j}||^{2} /2\sigma_{i}^{2})}{\sum_{k \neq i}{\text{exp}(-||\textbf{x}_{i} - \textbf{x}_{k}||^{2} /2\sigma_{i}^{2})}},& \text{if } i \neq j \\
            0, & \text{if } i = j
        \end{cases}
    \end{align}
    where $\sigma_i$ are the standard deviations. If we take two points $\textbf{x}_{i},\textbf{x}_{j}$, with $i\neq j$, then the values of $p_{i|j}$ and $p_{j|i}$ will be different. Therefore we define:
    \begin{align}
        p_{ij} = \frac{p_{j|i}+p_{i|j}}{2N}
    \end{align}
    \item Step 2: In the low dimensional space compute the similarity measures between datapoints, but instead of using a Gaussian distribution use a Student's t-distribution with one degree of freedom.
    \begin{align}
    q_{ij} = 
        \begin{cases}
            \frac{(1+||\textbf{y}_{i} - \textbf{y}_{j}||^{2})^{-1}}{\sum_{k \neq l}{(1+||\textbf{y}_{k} - \textbf{y}_{l}||^{2} )^{-1}}},& \text{if } i \neq j \\
            0, & \text{if } i = j
        \end{cases}
    \end{align}
    \item Step 3: Compute the Kullback-Liebler divergence between the two probability distributions $p_{ij}$ and $q_{ij}$. 
    \begin{align}
        D_{KL}(P||Q) = \sum_{i \neq j}{p_{ij}\log{\frac{p_{ij}}{q_{ij}}}}
    \end{align}
    This divergence is minimised with respect to the datapoints $\textbf{y}_{i}$ using gradient descent. 
\end{enumerate}

t-SNE has a hyper-parameter called perplexity, derived from $\sigma_i$, which is roughly a guess of the number of close neighbours each datapoint has and can have large effects on the resulting plot. Typical values of perplexity are in the range $[5,50]$. In our use of t-SNE we use perplexity equal to 48 which is the true number of webs in each class. 

%%%%%%%%%%

\end{document}